\documentclass[twocolumn,showpacs,preprintnumbers,amsmath,amssymb]{revtex4}
\usepackage{amssymb}
\usepackage{amsfonts}
\usepackage{mathrsfs}
\usepackage{amsmath}
\usepackage{graphicx}
\usepackage{dcolumn}
\usepackage{bm}
\usepackage{hyperref}

\begin{document}
\title{Single-photon-detection attack on the phase coding continuous variable quantum cryptography}
\author{Shi-Hai Sun, Mu-Sheng Jiang, Lin-Mei Liang\footnote{Email:nmliang@nudt.edu.cn}}
\affiliation{Department of Physics, National University of Defense
Technology, Changsha 410073, P.R.China}
\begin{abstract}
The phase coding quantum cryptographic scheme using the homodyne detection and weak coherent state [Hirano \emph{et al} 2003 \emph{Phys.Rev.A} \textbf{68} 042331] provides the simplest continuous variable quantum key distribution scheme from the experimental side. However, the inherent loss of practical system will not only increase the bit error rate (BER) but also affect the security of final key. In this paper, we propose a single-photon-detection attack, then the security of final key will be compromised in some parameter regimes. Our results show that the BER induced by Eve can be lower than the inherent BER induced by the loss of system in some parameter regimes. Furthermore, our attack gives the maximal communication distance of this scheme for given experimental parameters.
\end{abstract}


\maketitle
\section{\label{sec:Introduction}Introduction}
Quantum key distribution (QKD) admits two remote parties, known as Alice (the sender) and Bob (the receiver), to share secret key. The unconditional security of the key is guaranteed by the quantum mechanics, and any eavesdropper (Eve) in the quantum channel can be discovered by the legitimate parties. In the past years, the single-photon-based discrete variable QKD (DV-QKD), for example, the BB84 protocol \cite{BB84}, has been studied widely. The unconditional security of such scheme has been proved in theory for both the ideal system \cite{Lo99,Shor00} and the practical system \cite{GLLP04,Inamori07}. It has also been demonstrated with high repetition rate and long distance \cite{Stucki09,Zhang09,Eraerds10,Liu10}. Furthermore, there exists commercial products based on the DV-QKD \cite{Idq}. Although the DV-QKD has been well developed, it faces some inherent challenges within current technology, for example, the single photon source is not available nowadays and the efficiency of single photon detector is very low (about 20\%). As an alternative scheme for the DV-QKD, the continuous variable QKD (CV-QKD) is proposed \cite{Leverrier09,Braunstein05,Hirano03,Namiki03} and demonstrated \cite{Hirano03,Sheng10,Fossier09,Xu09}. Compared with the DV-QKD, the CV-QKD has two important advantages: one is that it uses the coherent state, instead of the single photon source, to encode the information, which can be obtained by using a commercial laser diode; the other one is that the homodyne detector used in the CV-QKD has very high efficiency (about 90\%).

In the CV-QKD, one interesting scheme is the application of the homodyne detection on the phase coding four-state protocol, which is proposed by Hirano \emph{et al.} \cite{Hirano03}. This scheme combines the phase modulator and the homodyne detection with a coherent pulse, thus it provides the simplest CV-QKD schemes from the experimental side. Although this scheme is secure for the ideal system \cite{Namiki03}, there exists inherent loss in the practical system (including the loss of channel and Bob's optical setups, and the imperfect efficiency of homodyne detector), which will not only raise the inherent bit error rate (BER) of system but also compromise the security of final key.

In this paper, we propose a single-photon-detection attack strategy to compromise, in some parameter regimes, the security of quantum cryptographic scheme using pulsed homodyne detection and weak coherent pulses \cite{Hirano03}. Our results show that the BER induced by our attack can be lower than the inherent BER of practical system in some parameter regime. Therefore, when our attack is taken into account, the secret key will be compromised. Furthermore, our results also give the maximal communication distance for given experimental parameters.

The paper is organized as following: In Sec.\ref{sec:scheme}, we review the quantum cryptographic scheme and then derive the bit error rate (BER) of Bob in the absence of Eve. In Sec.\ref{sec:attack}, we introduce Eve's strategy and derive the BER induced by Eve. In Sec.\ref{sec:discuss}, we first discuss one countermeasure to beat Eve's strategy. Then we discuss one open question. Finally, we give a brief summary of this paper in Sec.\ref{sec:summary}

\section{\label{sec:scheme}Quantum cryptography with pulsed homodyne detection}
In this section, we first review the quantum cryptographic scheme. Then we
derive the BER of Bob in the absence of Eve.

The protocol runs as follows \cite{Hirano03,Namiki03}. Alice randomly sends one of the four coherent states $\{|\alpha e^{ik\pi/2}\rangle\}$ to Bob, here $k=0,1,2,3$ and $\alpha$ is positive real number. The coherent state is the eigenstate of the annihilation operator $\widehat{a}$ of the light field. Then Bob randomly measures one of the two quadratures $\{\widehat{x}_1,\widehat{x}_2\}$. Here, $\widehat{x}_1+i\widehat{x}_2=\widehat{a}$, thus $[\widehat{x}_1,\widehat{x}_2]=i/2$. After the communication, Alice announces the basis used by her, when Bob uses the correct-basis, they keep the pulse, otherwise, they discard the pulse. Here we say a pulse is correct-basis means that Bob measures $\widehat{x}_1$ when Alice sends $|\pm \alpha\rangle$ and Bob measures $\widehat{x}_2$ when Alice sends $|\pm i \alpha\rangle$. For all the correct-basis pulses, Bob sets two threshold values $x_+>0$ and $x_-<0$ to judge the bit value of Alice. In the symmetric case, he can set $x_+=-x_-=x_0$. We assume Alice regards $\{|\alpha\rangle,|i\alpha\rangle\}$ as bit 0 and $\{|-\alpha\rangle,|-i\alpha\rangle\}$ as bit 1. Thus Bob's bit value is determined by
\begin{equation}
\text{(bit value)}=\begin{cases}0&\text{if $x>x_0$}\\1& \text{if $x<-x_0$}\\ \text{inconclusive} &\text{otherwise}, \end{cases}
\end{equation}
where $x$ is the measurement result of Bob's homodyne detection.

\begin{figure}
\scalebox{1}{\includegraphics[width=\columnwidth]{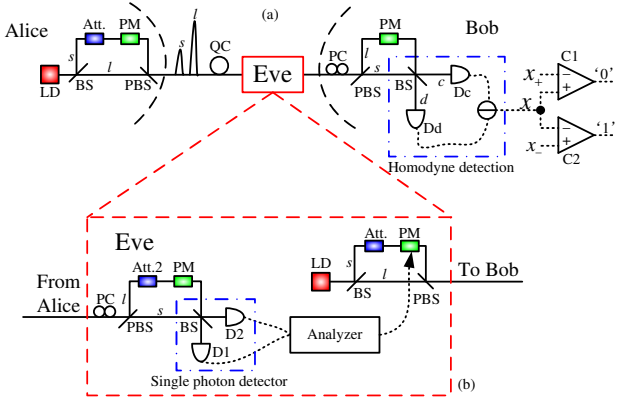}}
\caption{\label{fig:exp_arr}(Color online) The simple diagram of quantum cryptosystem with pulsed homodyne detector \cite{Hirano03} and Eve's experimental arrangement. LD: laser diode; BS: 50/50 beam splitter; PBS: polarization beam splitter; Att.: attenuator; PM: phase modulator; QC: quantum channel; PC: polarization controller; Dc and Dd: photodiode; C1 and C2: comparator; D1 and D2: single photon detectors (SPD). $x_+$ and $x_-$ are two threshold values set by Bob to judge the bit value of Alice. Generally speaking, he can set $x_+=-x_-=x_0$. \emph{s} and \emph{l} represent the signal pulse and local pulse respectively. Generally speaking, the intensity of local pulse will be larger than that of signal pulse, thus Att.2 is used by Eve to attenuate the intensity of local pulse to maximize the visibility of interference. Part(a) shows the general cryptosystem. Part(b) shows Eve's experimental arrangement. Eve intercepts the pulse from Alice and sends a faked pulse to Bob according to her measurement results.}
\end{figure}

The simple diagram of the experimental arrangement is shown in Fig.\ref{fig:exp_arr}(Part(a)). The pulse from Alice's laser is divided into two parties, one is the signal pulse (\emph{s}) and the other one is the local pulse (\emph{l}). Then Alice randomly and equally modulates one of four phase values $\{0,\pi/2,\pi,3\pi/2\}$ on the signal pulse with a phase modulator (PM). In order to ensure the security of key, Alice attenuates \emph{s} to a suitable level (about 1 photon/pulse). Note that, the intensity of \emph{l} is very large (about $10^6$ photon/pulse). When \emph{s} and \emph{l} arrive at Bob's zone, Bob measures $\widehat{x}_1$ or $\widehat{x}_2$ by randomly and equally modulating one of two phase values $\{0,\pi/2\}$ on the local pulse. Then the pulse will be detected by two photodiode (Dc and Dd). Finally, Bob uses two comparators to record the bit value.
In the following, we analyze the BER in the absence of Eve. Without loss of generality, we assume Alice sends $|\pm\alpha\rangle$ and Bob measures $\widehat{x}_1$. Thus the state received by Bob is given by
\begin{equation}
\rho_1=\frac{1}{2}(|\sqrt{\eta\mu_a}\rangle\langle\sqrt{\eta\mu_a}|+ |-\sqrt{\eta\mu_a}\rangle\langle-\sqrt{\eta\mu_a}|),
\end{equation}
where $\mu_a=|\alpha|^2$ is the intensity of signal state. $\eta=\eta_c\eta_{bob}$, here $\eta_c$ is the transmittance of channel, $\eta_{bob}$ is the transmittance of Bob's optical setups and the efficiency of homodyne detection. According to the measurement theory, the probability that the outcome $x_{\phi_b}$ is obtained by measuring $\widehat{x}_{\phi_b}=\widehat{x}_1\cos\phi_b+\widehat{x}_2\sin\phi_b$ of a coherent state $|\beta e^{i\phi_a}\rangle$ is given by \cite{Namiki03}
\begin{equation}\label{pro}
|\langle x_{\phi_b}|\beta e^{i\phi_a}\rangle|^2=\sqrt{\frac{2}{\pi}} \exp[-2(x_{\phi_b}-\beta \cos\phi)^2],
\end{equation}
where $\phi=\phi_b-\phi_a$. Thus the probability that Bob obtains a conclusive result is given by
\begin{equation}\label{post_absence}
\begin{split}
&P_{post}^{absence}\\=&\int_{-\infty}^{-x_0}dx_1\langle x_1|\rho_1|x_1\rangle + \int_{x_0}^{\infty}dx_1\langle x_1|\rho_1|x_1\rangle\\
=&\frac{1}{2}\{\text{erfc}[\sqrt{2}(x_0+\sqrt{\eta \mu_a})]+ \text{erfc}[\sqrt{2}(x_0-\sqrt{\eta \mu_a})]\},
\end{split}
\end{equation}
where $\text{erfc(x)}$ is the error function which is given by
\begin{equation}
\text{erfc}(x)=\frac{2}{\sqrt{\pi}}\int_{x}^{\infty}e^{-t^2}dt.
\end{equation}

Therefore, the inherent BER of Bob in the absence of Eve can be written as
\begin{equation}\label{BER_absence}
\begin{split}
&E_{bob}^{absence}\\=&\frac{1}{2P_{post}^{absence}}[\int_{-\infty}^{-x_0}|\langle x_1|\sqrt{\eta\mu_a}\rangle|^2 dx_1 +\int_{x_0}^{\infty}|\langle x_1|-\sqrt{\eta\mu_a}\rangle|^2 dx_1]\\
=&\frac{1}{2P_{post}^{absence}}\text{erfc}[\sqrt{2}(x_0+\sqrt{\eta \mu_a})].
\end{split}
\end{equation}

\begin{figure}
\scalebox{1}{\includegraphics[width=\columnwidth]{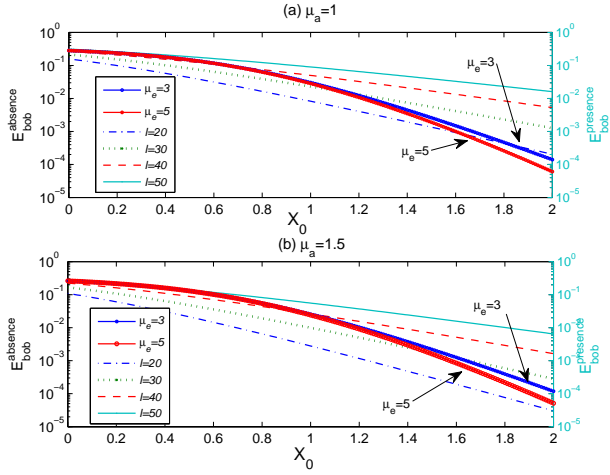}}
\caption{\label{fig:error_rate}(Color online) The bit error rate of Bob changes with the threshold value set by Bob. The dashed lines are the BER in the absence of Eve for different communication distance. The solid lines are the BER in the presence of Eve. $\mu_e$ is the intensity of pulse sent by Eve. $l$ is the distance of channel whose unit is \emph{km}. We assume the channel between Alice and Bob is fiber, thus $\eta_c=10^{-a l/10}$ ($a=0.21dB/km$ is the typical loss of fiber). In the simulations, we set $\eta_{bob}=0.6636$ according to the experimental result of Ref.\cite{Hirano03}. At the same time, we assume the SPD of Eve is perfect, which means $Y_0=0$ and $\epsilon=1$.}
\end{figure}

Eq.\ref{BER_absence} shows clearly that the inherent BER of system is determined by $x_0$, $\mu_a$ and $\eta$. Although Alice can choose $x_0$ and $\mu_a$ carefully to ensure that the system provides higher security, the loss of channel and homodyne detection is unavoidable. The BER for different communication distance is shown in Fig.\ref{fig:error_rate}. It shows clearly that the loss of channel raise the inherent BER of system quickly. Therefore, it may give some space to hide the existence of Eve. In fact, in next section, we will show that Eve can break the security of system in some parameter regime.

\section{\label{sec:attack}Single-photon-detection attack with SPD}
In this section, we first introduce Eve's attack strategy. Then we analyze the error rate induced by Eve's attack.

The experimental arrangement of Eve is shown in Fig.\ref{fig:exp_arr} (Part(b)). Eve first intercepts the pulse from Alice, then she randomly and equally modulates the local pulse with one of two phase ($0$ and $\pi/2$). Then local pulse and signal pulse will interfere at the BS and detected by two single photon detectors (SPDs). Note that the intensity of local pulse is much larger than that of signal pulse, thus Eve should attenuate the intensity of local pulse  to maximize the visibility of interference. If Eve obtains a valid result, she resends a faked state to Bob according to her measurement result, for example, if Eve modulates $0$, she resends $|\alpha\rangle$ (or $|-\alpha\rangle$) to Bob when D1 (or D2) clicks, if Eve modulates $\pi/2$, she resends $|i\alpha\rangle$ (or $|-i\alpha\rangle$) to Bob when D1 (or D2) clicks. Here valid result means that only one SPD clicks. If both of D1 and D2 click or neither of them click, Eve resends a vacuum state to Bob.

Obviously, Eve will introduce additional BER. However, we will show that, the BER introduced by Eve can be smaller than the BER introduced by the loss of practical cryptosystem in some parameter regime. Now, we analyze the BER introduced by Eve. Without loss of generality, we assume Alice sends $|\alpha\rangle$ and Bob measures $\widehat{x}_1$. When Eve modulates the local pulse with a phase $0$, the probabilities that D1 and D2 click are given by
\begin{equation}
\begin{split}
P_{D1}&=1-(1-Y_0)e^{-2\epsilon\mu_a}\\
P_{D2}&=Y_0,
\end{split}
\end{equation}
where we assume the interference of Eve is perfect. $Y_0$ is the dark count of Eve's SPD, $\epsilon$ is the efficiency of Eve's SPD. Thus, the probabilities that Eve resends $|\alpha\rangle$ and $|-\alpha\rangle$ are given by
\begin{equation}
\begin{split}
P_0^{co}&=P_{D1}(1-P_{D2})/2\\
P_1^{co}&=(1-P_{D1})P_{D2}/2,
\end{split}
\end{equation}
where the right side of the equation is divided by 2 since the probability that Eve modulates $0$ is 1/2. At the same time, Eve may modulate the local pulse with a phase $\pi/2$, then the probabilities that D1 and D2 click are given by
\begin{equation}
P'_{D1}=P'_{D2}=1-(1-Y_0)e^{-\epsilon\mu_a}\equiv P'_D.
\end{equation}
Thus, the probabilities that Eve resends $|i\alpha\rangle$ and $|-i\alpha\rangle$ are given by
\begin{equation}
P_0^{inco}=P_1^{inco}=P'_{D}(1-P'_{D})/2.
\end{equation}
Furthermore, the probability that Eve resends a vacuum state to Bob is given by
\begin{equation}
P_{vac}=[(1-P_{D1})(1-P_{D2})+P_{D1}P_{D2}+(1-P'_D)^2+(P'_D)^2]/2.
\end{equation}

Obviously, Eve's attack will disturb the original state sent by Alice, which will be transformed as
\begin{equation}\label{rho2rho}
\begin{split}
|\alpha\rangle\langle\alpha|\rightarrow& \rho'=P_0^{co}|\alpha_e\rangle\langle\alpha_e| +P_1^{co}|-\alpha_e\rangle\langle-\alpha_e|\\&+P_0^{inco}|i\alpha_e\rangle\langle i\alpha_e|+P_1^{inco}|-i\alpha_e\rangle\langle-i\alpha_e|\\&+P_{vac}|0\rangle\langle0|,
\end{split}
\end{equation}
where $|\alpha_e|^2$ is the intensity of pulse sent by Eve. As a result, the probability that Bob obtains conclusive result in the presence of Eve is given by
\begin{equation}\label{post_presence}
\begin{split}
&P_{post}^{presence}\\=&\int_{-\infty}^{-x_0}dx_1\langle x_1|\rho'|x_1\rangle + \int_{x_0}^{\infty}dx_1\langle x_1|\rho'|x_1\rangle\\
=&\frac{1}{2}(P_0^{co}+P_1^{co})\{\text{erfc}[\sqrt{2}(x_0+\sqrt{\mu_e})]+ \text{erfc}[\sqrt{2}(x_0-\sqrt{\mu_e})]\}\\
&+(P_0^{inco}+P_1^{inco}+P_{vac})\text{erfc}(\sqrt{2}x_0),
\end{split}
\end{equation}
where we set $\mu_e=\eta_{bob}|\alpha_e|^2$, since Eve can send a strong pulse to compensate the loss of Bob's optical setups and the efficiency of homodyne detector. And the BER in the presence of Eve can be written as
\begin{equation}\label{e_presence}
\begin{split}
&E_{bob}^{presence}\\
=&\frac{1}{P_{post}^{presence}}\int_{-\infty}^{-x_0}dx_1\langle x_1|\rho'|x_1\rangle\\
=&\frac{1}{2P_{post}^{presence}}\{P_0^{co}\text{erfc}[\sqrt{2}(x_0+\sqrt{\mu_e})]\\
&+P_1^{co}\text{erfc}[\sqrt{2}(x_0-\sqrt{\mu_e})]\\
&+(P_0^{inco}+P_1^{inco}+P_{vac})\text{erfc}(\sqrt{2}x_0)\}.
\end{split}
\end{equation}

The BER in the presence of Eve is shown in Fig.\ref{fig:error_rate}. It shows clearly that the BER induced by Eve can be lower than the BER induced by the loss of system in some parameter regime. For example, when $\mu_a=1$ and $l=30 km$, $E_{bob}^{presence}$ is smaller than $E_{bob}^{absence}$ for the threshold value that $x_0>1.12$. Thus, in these parameters regime, the loss of system will leave some space for Eve to spy the secret key and the security of final key will be compromised. In other words, the legitimate parties must set their experimental parameters carefully to remove the existence of Eve.

Here we remark that Fig.\ref{fig:error_rate} gives the maximal threshold value that can be set by Bob for a given communication distance, then it gives the minimal inherent BER of system. For example, when $l=30 km$, the maximal threshold value is 1.12 and 1.47 for $\mu_a=1$ and $\mu_a=1.5$ respectively, which corresponds to the minimal inherent BER 1.65\% and 0.19\% respectively. Furthermore, the simulations show that when $l=50 km$, $E_{bob}^{presence}$ is always smaller than $E_{bob}^{absence}$, even Bob sets $x_0=0$. Thus our attack also gives the maximal communication distance of the cryptographic scheme for given experimental parameters.

\section{\label{sec:discuss}Discussion}

\subsection{Countermeasure}
In Sec.\ref{sec:attack}, we have shown that the BER induced by our attack can be lower than the BER induced by the inherent loss of cryptosystem in some parameter regime. Thus the legitimate parties must set their experimental parameters carefully to remove the existence of Eve. However, these limited parameters regime will affect the performance of the cryptosystem. In this section, we discuss one countermeasure to show that the legitimate parties can discover the existence of Eve by reconstructing the probability density of each state sent by Alice.

Without loss of generality, we also assume that Alice sends $|\alpha\rangle$. Thus when Eve is present, the original state will be transformed as $\rho'$ which is given by Eq.\ref{rho2rho}. It is easy to derive the probability density of Bob's measurement result $x$. When Bob uses the correct basis, the probability density in the absence and presence of Eve are given by
\begin{equation}
\begin{split}
P_{co}^{absence}&=\langle x_1|\sqrt{\eta\mu_a} \rangle\langle\sqrt{\eta\mu_a}|x_1\rangle\\
P_{co}^{presence}&=\langle x_1|\rho'|x_1\rangle,
\end{split}
\end{equation}
and when Bob uses the wrong basis, the probability density are given by
\begin{equation}
\begin{split}
P_{inco}^{absence}&=\langle x_2|\sqrt{\eta\mu_a} \rangle\langle\sqrt{\eta\mu_a}|x_2\rangle\\
P_{inco}^{presence}&=\langle x_2|\rho'|x_2\rangle.
\end{split}
\end{equation}

\begin{figure}
\scalebox{1}{\includegraphics[width=\columnwidth]{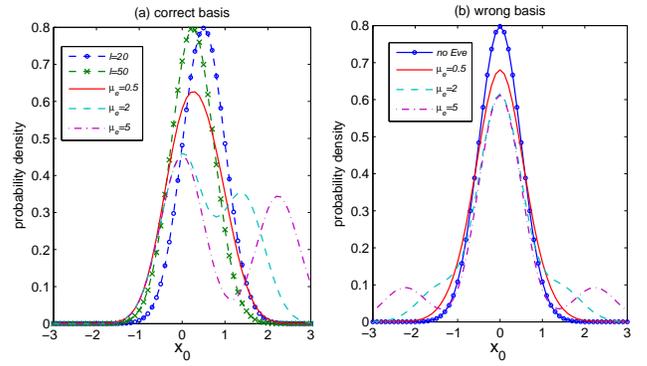}}
\caption{\label{fig:pro_dis}(Color online) The probability density of quadrature $x$ measured by Bob. The figure shows the probability density when Alice sends $|\alpha\rangle$, thus the correct basis means that Bob measures $\widehat{x}_1$ and the wrong basis means that Bob measures $\widehat{x}_2$. The dashed lines are drawn in the absence of Eve and the solid lines are obtained in the presence of Eve. Here we set $\mu_a=1$. Other parameters used in the simulations are the same as that of Fig.\ref{fig:error_rate}. }
\end{figure}

The probability density is shown in Fig.\ref{fig:pro_dis}. It shows clearly that, when Eve is absent, the probability density of $x$ still follows the Gauss distribution, and the loss of system only reduces the average value. But Eve's attack will change the probability density of Bob's measurement (see the solid lines of part(a) in Fig.\ref{fig:pro_dis}), since she may send wrong state to Bob. Thus Bob can discover the existence of Eve by reconstructing the probability density of his measurement result.

Although Eve's attack will affect the probability density of $x$ when Bob uses the correct basis, the difference is small when Bob uses the wrong basis (see part(b) of Fig.\ref{fig:pro_dis}). The simulation shows that $x$ still follows the Gauss distribution when Bob chooses the wrong basis, even Eve is present. Thus, from the experimental side, it is hard to know the difference is caused by Eve or the statistical fluctuation of the measurement result. In other words, the legitimate parties only need to reconstruct the probability density of $x$ for the case that Bob uses the correct basis, and they can discard the case that Bob uses the wrong basis.

\subsection{Compare with previous attack}
\begin{figure}
\scalebox{1}{\includegraphics[width=\columnwidth]{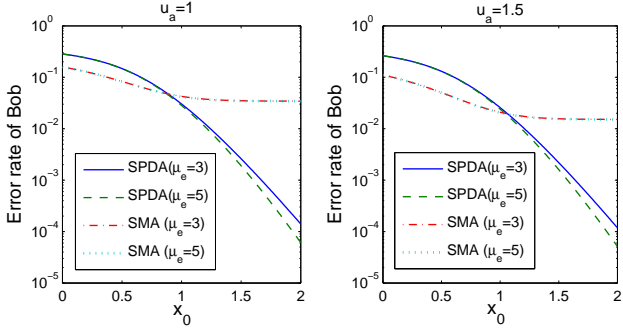}}
\caption{\label{fig:compare}(Color online)The error rate of Bob in the presence of Eve. SPDA represents the single-photon-detection attack, and SMA is the simultaneous measurement attack which is proposed and analyzed by Namiki and Hirano in the Ref.\cite{Namiki03}. In the simulations, we assume Eve's detection system is perfect, which means that both the SPDs and homodyne detectors of Eve are perfect.}
\end{figure}

In this paper, we propose a new method to attack the phase coding continuous variable quantum cryptography. According to the analysis described above, it is known that the security of final key will be compromised in some parameter regimes, when our attack is taken into account. In fact, our attack can be classified as an intercept-and-resend attack, but Eve uses SPDs to read out Alice's information, instead of using a homodyne detection system. In order to show the advantage of our attack, we compare our attack with the simultaneous measurement attack (SMA), which is analyzed by Namiki and Hirano in the Ref.\cite{Namiki03}. In SMA, Eve equivalently splits the signal pulse into two parts with a 50:50 beam splitter, and measures $\widehat{x}_1$ of one part and $\widehat{x}_2$ of the other part. Her measurement results are noted as $(x_1,x_2)$ for each signal pulse. Then she resends a new coherent state to Bob according to the inequality between $x_1$ and $x_2$, which is given by
\begin{equation}
\begin{matrix}
|\alpha_e\rangle &\text{if} &x_1\geq|x_2|\\
|-\alpha_e\rangle &\text{if} &-x_1\geq|x_2|\\
|i\alpha_e\rangle &\text{if} &x_2>|x_1|\\
|-i\alpha_e\rangle &\text{if} &-x_2>|x_1|,\\
\end{matrix}
\end{equation}
where $|\alpha_e|^2$ is the intensity of pulse sent by Eve. Therefore, when Eve loads th SMA, Eq.\eqref{rho2rho} will be rewritten as \cite{Namiki03}
\begin{equation}\label{rho2rhoSMA}
\begin{split}
|\alpha\rangle\langle\alpha|\rightarrow& \rho''=P_+|\alpha_e\rangle\langle\alpha_e| +P_-|-\alpha_e\rangle\langle-\alpha_e|\\&+P_\bot(|i\alpha_e\rangle\langle i\alpha_e|+|-i\alpha_e\rangle\langle-i\alpha_e|),
\end{split}
\end{equation}
where
\begin{equation}
\begin{split}
P_+&=\int_{x_1\geq|x_2|}Q(x_1,x_2)dx_1dx_2\\
P_-&=\int_{-x_1\geq|x_2|}Q(x_1,x_2)dx_1dx_2\\
P_\bot&=\int_{x_2>|x_1|}Q(x_1,x_2)dx_1dx_2,\\
\end{split}
\end{equation}
and
\begin{equation}
Q(x_1,x_2)=\frac{2}{\pi}\exp[-2(x_1-\sqrt{\mu_a/2})^2-2x_2^2].
\end{equation}

Then we can estimate the error rate of Bob in the presence of Eve by using Eq.\eqref{post_presence} and Eq.\eqref{e_presence}, which is shown in Fig.\ref{fig:compare}. It clearly shows that, in some parameter regimes, the error rate of Bob in the presence of Eve induced by our attack can be much lower than that of SMA.

\begin{table}
\caption{\label{tab:Table1} The probability that Eve resends the four new coherent states and vacuum state to Bob for our attack and SMA respectively. Here we assume that Alice sends $|\alpha\rangle\langle\alpha|$. In the simulation, we set $\mu_a=1$ and $\mu_e=3$, and assume both the SPD and homodyne detector of Eve are perfect.}
\tabcolsep0.05in
\doublerulesep 2pt
\begin{tabular}{cccccc}
\hline\hline
&$|\alpha\rangle\langle\alpha|$ &$|-\alpha\rangle\langle-\alpha|$ &$|i\alpha\rangle\langle i\alpha|$ &$|-i\alpha\rangle\langle-i\alpha|$ &$|0\rangle\langle0|$\\
\hline
SPDA &0.4323 &0 &0.1163 &0.1163 &0.3351\\
\hline
SMA &0.7079 &0.0252 &0.1334 &0.1334 &0\\
\hline\hline
\end{tabular}
\end{table}

Here we give a qualitative explanation about this result. Without loss of generalization, we assume that Alice sends $|\alpha\rangle$, and both the SPDs and homodyne detector of Eve are perfect. Then it is easy to check that, when Eve chooses the same basis with Alice, she can determinately obtain the correct result ($|\alpha\rangle$) if she uses the SPDs to measure Alice's signal state, but she will obtain the wrong result ($|-\alpha\rangle$) with nonzero probability if she uses the homodyne detector. At the same time, when Eve chooses the incorrect basis, she will obtain $|i\alpha\rangle$ and $|-i\alpha\rangle$ with equivalent probability. Of course, when Eve uses the SPDs, she may resend the vacuum state $|0\rangle$ to Bob since there exists empty pulses in coherent states. Table \ref{tab:Table1} shows the typical probability that Eve resends the four new coherent states and the vacuum state to Bob for our attack and SMA respectively. Therefore, in view of Bob (note that Bob will use the homodyne detector to measure the signal state from Alice), the probability that he obtains the result $x$ is given by
\begin{equation}\label{P_bob}
\begin{split}
P(x)&=\sqrt{2/\pi}\{P_{\alpha}\exp[-2(x-\alpha_e)^2]\\
&+ P_{-\alpha}\exp[-2(x+\alpha_e)^2]+ P_{0}\exp[-2x^2]\},
\end{split}
\end{equation}
where $P_{\pm\alpha}$is the probability that Eve sends $|\pm\alpha\rangle$ to Bob. $P_{0}$ is the total probability that Eve sends $|\pm i\alpha\rangle$ and $|0\rangle$ to Bob. Thus $P_{\alpha}=P_{0}^{co}$, $P_{-\alpha}=P_{1}^{co}=0$, $P_{0}=P_{0}^{inco}+P_{1}^{inco}+P_{vac}$ for our attack, and $P_{\alpha}=P_{+}$, $P_{-\alpha}=P_{-}$, $P_{0}=2P_{\bot}$ for SMA.
Note that the first term of Eq.\eqref{P_bob} is the correct term which corresponds to the case that Eve sends the correct coherent state $|\alpha\rangle$ to Bob, but the second and third term are incorrect terms. However, the third term is a gauss distribution with a center value of 0, and will decrease quickly with the $x$. Thus when the threshold value ($x_0$) set by Bob is large enough, the equivalent state received by Bob for our attack and SMA can be written as
\begin{equation}
\begin{split}
\rho'&\rightarrow |\alpha_{e}\rangle\langle\alpha_{e}|\\
\rho''&\rightarrow P_{\alpha}|\alpha_{e}\rangle\langle\alpha_{e}|+P_{-\alpha}|-\alpha_{e}\rangle \langle-\alpha_{e}|.
\end{split}
\end{equation}
Here the equivalent state means that it has the same probability distribution as Eq.\eqref{P_bob}. Therefore, our attack can perform better than SMA when the threshold value ($x_0$) set by Bob is larger than a given value. However, note that the probability $P_0$ of our attack is larger than that of SMA, thus when $x_0$ is small, SMA will perform better than our attack, since the third term can not be ignored at this time.

\subsection{Open question}
In this paper, we show that Eve can use the SPD to spy the secret key, then the security of quantum cryptographic scheme using homodyne detection and weak coherent state will be compromised in some parameter regimes. But the legitimate parties can discover the existence of Eve by reconstructing the probability density of state sent by Alice. In fact, this work is a consecutive work of our previous work \cite{Sun12}, in which we proposed a partially random phase (PRP) attack and show that the homodyne detection can by used to break the security of BB84 protocol based on SPD if the phase of source is just partially randomized. But the PRP attack can be beaten by the decoy state method \cite{Hwang03,Lo05,Wang05,Ma05}, which can be considered as that the legitimate parties reconstruct the character of channel (for example, the yield and error rate of each \emph{n}-photon pulse). Furthermore, note that, in the practical QKD system, the legitimate parties may just use part of the dimension of the photon pulse sent by Alice, thus other dimensions of the photon pulse may leave some loopholes for Eve to spy the secret key. For example, in the phase coding QKD system, only phase dimension is used, but the photon number dimension of each pulse is not used which will cause the photon-number-splitting (PNS) attack \cite{Huttner95,Brassard00}. But the PNS attack can be beaten by reconstructing the character of channel with the decoy state method. All of these attacks may imply that, if the legitimate parties reconstruct the character of photon pulse in the practical QKD system, they can beat Eve's attack strategy based on the imperfection of photon pulse used by them.

\section{\label{sec:summary}Summary}
The phase coding quantum cryptographic scheme based on the pulsed homodyne detection and weak coherent state \cite{Hirano03} provides the simplest CV-QKD schemes from the experimental side, since only the phase modulator and homodyne detector are needed. However, there exists inherent loss in the practical system (including the loss of channel and Bob's optical setups, and the imperfect efficiency of homodyne detector), which will not only raise the inherent BER of system but also compromise the security of final key in some parameter regime.

In this paper, we propose an attack to compromise, in some parameter regimes, the security of final key generated by this scheme. Our results show that the BER induced by Eve can be lower than the BER induced by the loss of system in some parameter regime. Thus, when our attack is taken into account, the security of final key will be compromised. Furthermore, our attack limits the maximal threshold value that can be set by Bob for given communication distance, thus it also give the minimal inherent BER of system. Note that, for a given communication distance, if the BER induced by Eve is always lower than the inherent BER induced by loss of system, no secret key can be generated in this distance. In other words, our attack also give the maximal communication distance of this scheme for given experimental parameters.

\section{ACKNOWLEDGEMENT}
This work is supported by the National Natural Science Foundation of China, Grant No. 61072071. Lin-Mei Liang is supported by Program for NCET. S.H. Sun is supported by the Hunan Provincial Innovation Foundation for Postgraduates, Grant No. CX2010B007, and the Fund of Innovation, Graduate School of NUDT, Grant No. B100203.


\end{document}